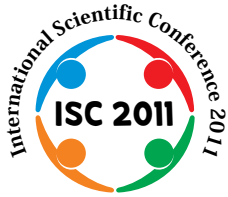

**Proceeding of the International Conference on Advance Science, Engineering and Information Technology 2011**

Hotel Equatorial Bangi-Putrajaya, Malaysia, 14 - 15 January 2011

ISBN 978-983-42366-4-9

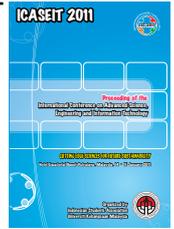# Metamorphic Virus Detection in Portable Executables Using Opcodes Statistical Feature

Babak Bashari Rad[#], Maslin Masrom[*]

[#] Faculty of Computer Science and Information Technology, University Technology of Malaysia
*Jalan Semarak UTM KL Campus, Kuala Lumpur, 54100, Malaysia*
*Tel.:+60326934844, E-mail: babak.basharirad@hotmail.com*

[*] UTM Razak School of Engineering and Technology, University Technology of Malaysia
*Jalan Semarak UTM KL Campus, Kuala Lumpur, 54100, Malaysia*
*Tel.:+60326934844, E-mail: maslin@ic.utm.my**Abstract*— Metamorphic viruses engage different mutation techniques to escape from string signature based scanning. They try to change their code in new offspring so that the variants appear non-similar and have no common sequences of string as signature. However, all versions of a metamorphic virus have similar task and performance. This obfuscation process helps to keep them safe from the string based signature detection. In this study, we make use of instructions statistical features to compare the similarity of two hosted files probably occupied by two mutated forms of a specific metamorphic virus. The introduced solution in this paper is relied on static analysis and employs the frequency histogram of machine opcodes in different instances of obfuscated viruses. We use Minkowski-form histogram distance measurements in order to check the likeness of portable executables (PE). The purpose of this research is to present an idea that for a number of special obfuscation approaches the presented solution can be used to identify morphed copies of a file. Thus, it can be applied by antivirus scanner to recognize different versions of a metamorphic virus.

*Keywords*— Metamorphic Virus, Obfuscation Techniques, Virus Detection, Opcode Frequency Histogram## I. INTRODUCTION

Metamorphic virus problem is still an inconvenience subject in computer virology, in recent years. Nearly all current anti-virus detectors are relied on syntactic properties of viruses. They usually scan sequences of bytes in machine code of executable files to match the signature of the virus. It makes the anti-virus scanner at risk of the smart virus attacks which are employing mutation techniques more and more [1].

In recent years, antivirus products tend to use of semantic features to defeat the metamorphic viruses' obfuscation. The most essential problem in regard to semantic approaches is need for a lot of preliminary tasks. It requires plenty of time for analysing and producing a suitable semantic signature. Furthermore, semantic methods are not practical in on-the-fly scanning [2]. Also, a method that needs a great deal of time to analyse and detect a mutated copy of a virus is not logical, though it may be enough clever to detect the virus precisely.

Conversely, syntactic signature strategy needs continually updated records of signatures. So, to keep a trustful database, many experts should spend time to extract string signatures and update database with as the most truthful data as possible [3]. However, all these endeavours can be simply frustrated by metamorphism techniques.

Metamorphic virus authors make use of various morphing techniques to escape from signature-based analysis. Some obfuscation skills frequently used by them are 1) garbage code insertion, 2) register usage exchange, 3) instruction replacement, 4) instruction permutation and 5) Code Transposition [2], [4], [5], [6] and [7].

In this paper, we use the histogram of instructions opcodes as a statistical feature to check the similarity of the executables. We try to recognize is a given file a morphed copy of another one. We think that while obfuscation functions change the visual structure of morphed variants of a virus, but many essential common features with these variants will be still remained, which contain the basic instructions and present their related functions. In other words, mutation engines cannot obliterate statistical similarity properties of two codes that are similar in behaviour and performance.

403

Next section review some related works has been done, previously. In section 3, we review some of the most common obfuscation rules generally used by metamorphic virus engines. Our method is introduced in section 4, and experiments and result are presented in section 5. Lastly, we give the conclusion and some suggestions for future researches.

## II. RELATED WORKS

In [8], Szor and Ferrie introduced a valuable definition of metamorphic viruses and evolution of the code. They also introduced some basic metamorphic virus detection methods, in general, following with many useful examples.

Konstantinou, in his technical report [5], gave a comprehensive and detailed explanation for metamorphic viruses, obfuscation techniques and other advanced skills normally used by them. Then, he discusses about metamorphic virus detection methods, briefly.

The method introduced in [1] is based on this concept that properties of malwares are positioned in their semantics. Preda et al in this paper recommended a semantics-based structure for malware detectors. Their approach uses trace semantics to distinguish the behaviours of malware while the program code is being inspected for infection.

A helpful explanation of computer virus strategies and detection methods is accessible in [9] by authors. They explained static and dynamic detection approaches, mechanism of metamorphic virus engine and open problems in computer anti-virus technologies.

In [2], Karnik et al presented a method based on frequency of instructions using cosine similarity analysis to detect obfuscated viruses.

Webster and Malcolm in [10] introduced an approach towards metamorphic computer virus detection by an algebraic specification of the IA-32 assembly programming language. Their proposed method based on a specification in OBJ of the IA-32 instructions.

## III. OBFUSCATION TECHNIQUES

As mentioned in previous section, metamorphic viruses utilize different techniques to defeat string signature based detection. In fact, metamorphic virus is able reprogram itself to challenge deeper static analysis [9].

In following, we review some of popular obfuscation techniques with examples of morphed codes, to understand how obfuscation may change the sequence of bytes in an executable to neutralize scanning.

### A. Garbage Code Insertion

The simplest technique used by metamorphic engine to change the byte sequence of viral code is garbage code (or dead code) insertion. Inserted instruction has no effect on function of the code. There are different kinds of garbage code insertion.

However, more mixed and complicated techniques of different types of garbage code insertion can be used in metamorphic viruses.

A more detailed descriptions and examples can be found in [6] and [11].

### B. Register/Variable Usage Exchange

Usage of different registers or memory variables is another simple transformation method that metamorphic engines use it to mutate their code. This technique attempt to evade the string signature based detection as well, by changing similar bytes in various generations. In December 1998, Win95.Regswap utilized it to create different variants of the virus. It is clear that it does not influence on the function of the code, but the sequence of binaries will alter. More detailed information and examples can be found in [8] and [11].

The combination of this technique with other methods such as dead code insertion can make new generations enough difficult to detect and make the syntax signature based detection entirely unusable.

### C. Instruction Replacement

This obfuscation method actually substitutes some instructions with their equivalent instructions in newer copies. Sometimes, programmers can perform an action in different ways of coding. For example, to assign 0 to register eax, following codes are possible:

mov eax, 0
xor eax, eax
and eax, 0
sub eax, eax

Therefore, this is a great opportunity for the virus programmers to utilize this possibility in metamorphic engines. This method is like using different synonyms in human language [2].

W95.Bistro virus uses this technique to transform its code. Szor in [4] gives a detailed investigation and examples for this virus.

### D. Instruction Permutation

In some pieces of code, it is possible to change the sequence of instructions with no disturbing the execution. Byte strings in different versions of the code will appear unlike via this disordering technique.

If there is no dependency among some instructions, they can be reordered. Consider the following instructions:
op1       Reg1, Reg2
op2       Reg3, Reg4

If the below conditions are satisfied, these two instructions can be substituted [7]:
1- Reg1 is not equal to Reg2
2- Reg1 is not equal to Reg4
3- Reg2 is not equal to Reg3

### E. Code Transposition

This technique modifies the structure of the program in form of physically reordering of the program codes, while preserving the execution order or flow of the program running using conditional jumps or unconditional branches. It may be done at the level of instructions or modules.

An example of such code structure modification utilized by *Zperm* virus is presented by Szor and Ferrie in [8], in more detail.

## IV. PROPOSED METHODOLOGY

In some mutation techniques, especially for register/variable exchange, or instruction permutation or even in some lower level of code transposition, the number of same instructions is nearly equivalent in various copies of morphed viruses. Our proposed solution deals with the frequencies of opcodes used in variants as a feature and calculates the dissimilarity between two files. We expect that if the obfuscation engine employs some particular morphing techniques, the frequencies of identical instructions are almost equal.

In addition, to achieve a better comparison, we breakdown the files into their building subroutines and compare two files according to their function blocks.

We make an instruction frequency histogram for each code block or subroutine. Then we can evaluate dissimilarity of two blocks by measure the distance between their histograms. It can be done by different histogram distance measurements techniques introduced in data mining techniques. In fact, each subroutine of a program is presented by a histogram of the contained instructions as a feature, in form of a vector, which the length of vector is equal to the number of total instructions of the machine. If we were able to compare the histograms of the building blocks of two programs, then we will be able to calculate the dissimilarity between two programs using this feature.

Therefore, if dissimilarity value between two programs is less than a specific threshold we can conclude these two programs are morphed versions of each other. By this way, we can classify different variants of a metamorphic virus, which use some special types of obfuscation techniques. It is significant to mention that we can use this approach in the cases which the applied obfuscation has no effects or small changes on the frequencies of instructions

### A. Data Structure and Algorithm

To measure the dissimilarity between two executable files, we follow two general steps. In first step, is a pre-process, we prepare our input data in form of histograms as features. In the second step, which is a comparison process, we evaluate the dissimilarity of a pair of programs by comparison of their histograms.

In pre-process section, first, we disassemble executable files using IDA Pro 4.9 [12] and create assembly code files. Then, we analyze each assembly program and extract all procedures inside and save them as separate files. In next step, we create a set of histograms represent the frequencies of instructions within the procedures for each file. As a result, for each program, we will have a set of histograms, each one for a sub-procedure. Fig. 1 shows the process of program disassembly and breakdown into building subroutine blocks.

In the second step, we compare two programs by the use of their sets of histograms. Our comparison method is similar to that proposed in [2] with some changes to improve the algorithm. The comparison algorithm works as follow:

We have a set of histogram for each program. Each histogram driven for a subroutine inside the program and includes the frequency of each instruction in the subroutine.

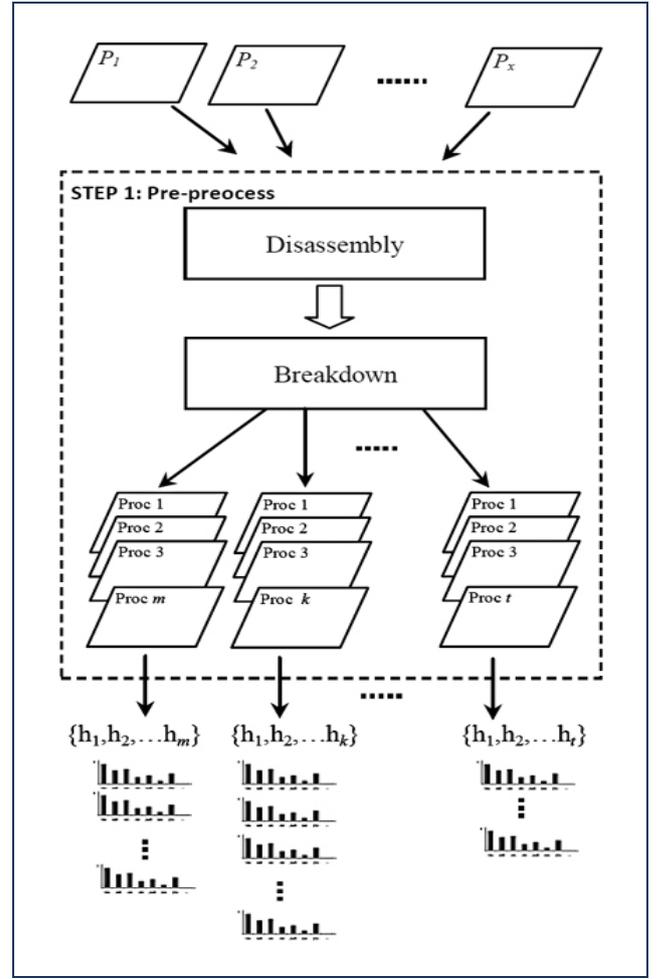

Fig. 1 Disassembly, breakdown and feature extraction

Given two programs $P1$ and $P2$, containing $m$ and $k$ subroutines, respectively. Therefore, $P1$ has $m$ histograms and $P2$ has $k$ histograms. Each histogram of $P1$ is compared with all histograms of $P2$. According to our distance metric introduced in next section; distance value for each comparison will be calculated. More precisely, histogram $h_i$ from $P1$ is compared with all histograms $h_j$, $1 \leqslant j \leqslant k$, from $P2$. That pair of histograms which has the minimum distance considered as the most similar histograms and consequently, we can consider their corresponding subroutines as mutated versions of each other. We save the minimum distance value for subroutine $i$. Once all histograms of $P1$ compared with all histograms of $P2$, we have a vector of length $m$, which contains the minimum distance values. We can use the average of this vector as the total distance value of $P1$ and $P2$. It is important to take notice that this distance value is not symmetric. It means $distance(P1, P2)$ is not equal to $distance(P2, P1)$. Hence, to get a more precise result, we define the distance of $P1$ and $P2$ as following:

$$d\{P1,P2\} = \frac{d(P1,P2) + d(P2,P1)}{2} \qquad (1)$$

Fig. 2 shows the process of comparison between two programs $P1$ and $P2$, briefly.

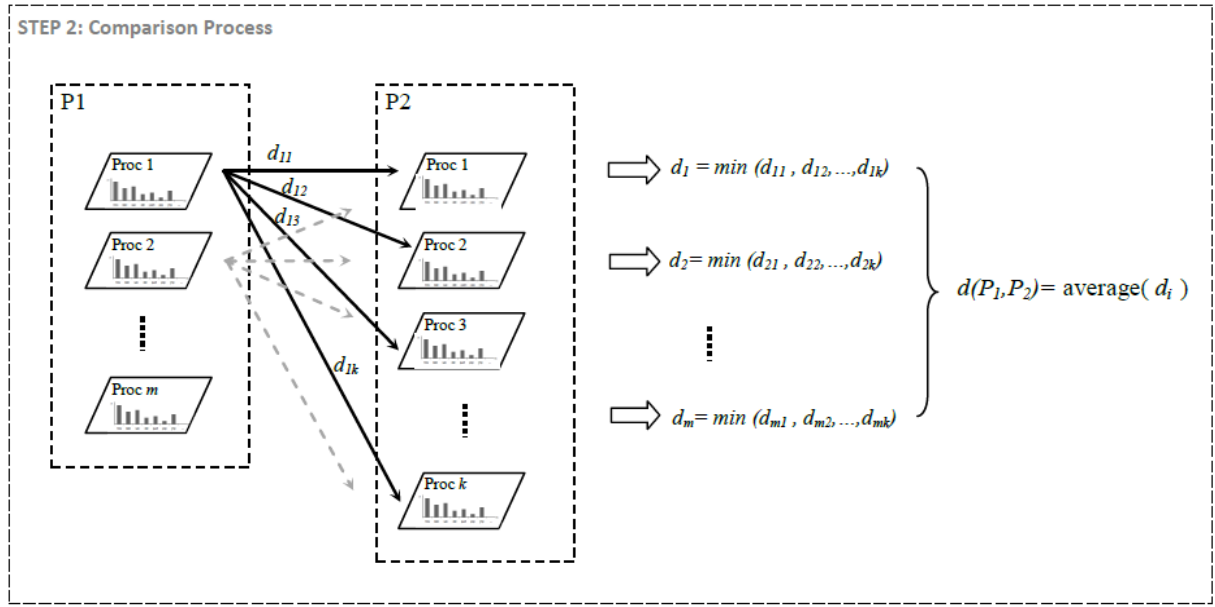

Fig. 2 Distance Calculation between two programs *P1* and *P2*

## B. Dissimilarity Metric

There are various metric methods for measuring histograms distance. The first class of dissimilarity measurement is based on Minkowski-form distance metric [13]. Consider two vectors of size $n$, $X = (x_1, x_2, ..., x_n)$ and $Y = (y_1, y_2, ..., y_n)$, then the Minkowski-form distance between two vectors X and Y is calculated as:

$$d^r_{X,Y} = \sum_{i=1}^{n} |x_i - y_y|^r \qquad (2)$$

Two popular histogram distance measurements are Manhattan and Euclidean form distance, Minkowski-form metrics with $r = 1$ and $r = 2$, as following:

$$d_{X,Y} = \sum_{i=1}^{n} |x_i - y_i| \qquad (3)$$

$$d^2_{X,Y} = \sum_{i=1}^{n} (x_i - y_i)^2 \qquad (4)$$

As shown in Fig. 3, histogram dissimilarity measures based on Minkowski-form compare only the parallel elements. It is appropriate for our case that the instructions opcodes are independent variables.

In addition, because we are going to test different kinds of programs in our data set, as we explained in 5.2, to obtain a common threshold for classification, we normalized the histograms before we begin to calculate the distance values as following:

$$X = \frac{X}{\sum_{i=1}^{m} x_i} \qquad (5)$$

Histogram normalization supports us to find a program independent threshold for our data set.

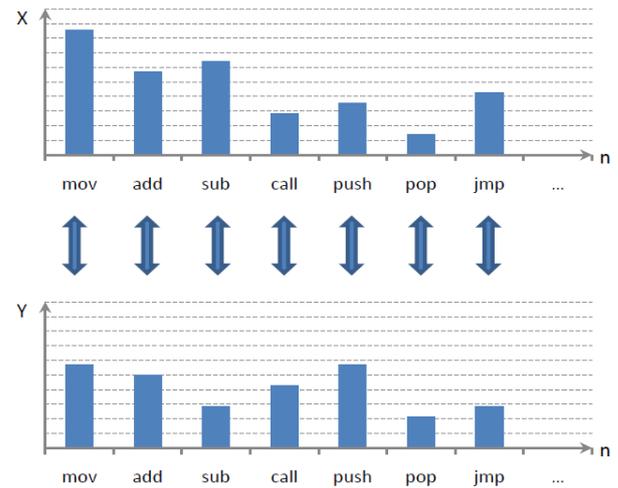

Fig. 3 Minkowski-form distance metrics compare only identical bins between two histograms.

## V. EXPERIMENTS AND FINDINGS

We used MathWorks MATLAB R2008a [14] to carry out our experiments include of preparing the data structure and implementing the comparison algorithm, and distance calculation.

### A. Data Set

In our test data collection, we used different obfuscated versions of two famous metamorphic viruses, Win32/Evol and Win32/Evul, retrieved from [15] and a number of randomly chosen benign programs. Viruses and legal programs used in the experiment listed in Table I.

An in detail investigation and analysis of Win32.Evol is given in [16].

TABLE I
LIST OF EXPERIMENT FILES

| Virus/Program | Description/Application |
|---|---|
| Win32.Evol.a/b/c | Versions of virus Evol |
| Win32.Evul. a/b/c/d/e/f/g/h | Versions of virus Evul |
| Lib.exe | Microsoft Visual Studio 9.0 |
| Write.exe | Windows 7 WordPad |
| Wordconv.exe | Microsoft office 2007 |
| Help.exe | Windows 7 command line help |
| Find.exe | Windows 7 command line find |
| Drvins32.exe | Kaspersky Antivirus 2010 |
| Perlglob.exe | Mathworks MatLab R2008a |
| logoUI.exe | Windows 7 |

*B. Results and Discussion*

Table II shows the comparison results for each pair of two files. The lower triangle (yellow cells) in table, gives the distance values according to Manhattan distance or Minkowski-form with *r=1*, and the upper triangle (blue cells) is the Euclidean distance values or Minkowski-form with *r=2*. Highlighted cells indicate values less than chosen thresholds. Some of false negatives and false positives are specified in the table.

The lower values denote that two programs are more similar. If the distance value is lower than a specified threshold, then we can deduce those two programs are obfuscated copies. So, deciding on the suitable threshold value is important. Choosing a low threshold value may cause to produce false negative and a very high threshold can increase false positives. We believe, generally, threshold value must be considered according to the case. For different viruses with not the same mutation engines and morphing degree, threshold values may be chosen absolutely different.

In the case of our study, distance values between each pairs of three versions of *Evol* virus are equal to zero. Anyway, if we use some other versions of this virus, maybe we have to choose a higher threshold to classify the other morphed instances of this virus. It can be seen for *Evul* virus variants. If we tend to remove the false positives, we have to choose a threshold less than **0.062**, for example, in the case of Euclidean distance.

As it is shown in the Table II, in both distance metrics, three instances *Evul.d*, *Evul.f* and *Evul.g* have higher distance values. After analysing *Evul* Virus deeply, we find that in these three variants, virus inserts frequently *nop* instruction as dead code. For this reason, histograms of these versions are disparate and distance values show more dissimilarity.

TABLE II

MINKOWSKI-FORM DISTANCE VALUES AND CLASSIFICATION

THRESHOLDS: 0.832 FOR MANHATTAN DISTANCE (r=1) AND 0.186 FOR EUCLIDEAN DISTANCE (r=2)

| | Win32-Evol-a | Win32-Evol-b | Win32-Evol-c | Win32-Evul-8192-a | Win32-Evul-8192-b | Win32-Evul-8192-c | Win32-Evul-8192-d | Win32-Evul-8192-e | Win32-Evul-8192-f | Win32-Evul-8192-g | Win32-Evul-8192-h | lib | write | wordconv | help | find | drvins32 | perlglob | LogonUI-til |
|---|---|---|---|---|---|---|---|---|---|---|---|---|---|---|---|---|---|---|---|
| Win32-Evol-a | 0.000 | 0.000 | 0.000 | 0.426 | 0.393 | 0.395 | 0.538 | 0.426 | 0.538 | 0.535 | 0.426 | 0.304 | 0.317 | 0.230 | 0.303 | 0.259 | 0.368 | 0.593 | 0.259 |
| Win32-Evol-b | 0.000 | 0.000 | 0.000 | 0.425 | 0.392 | 0.395 | 0.537 | 0.425 | 0.537 | 0.534 | 0.425 | 0.303 | 0.315 | 0.230 | 0.302 | 0.258 | 0.367 | 0.592 | 0.258 |
| Win32-Evol-c | 0.000 | 0.000 | 0.000 | 0.425 | 0.391 | 0.394 | 0.536 | 0.425 | 0.536 | 0.534 | 0.425 | 0.303 | 0.315 | 0.230 | 0.302 | 0.258 | 0.367 | 0.591 | 0.258 |
| Win32-Evul-8192-a | 1.675 | 1.674 | 1.674 | 0.000 | 0.036 | 0.032 | 0.149 | 0.000 | 0.149 | 0.215 | 0.004 | 0.523 | 0.543 | 0.340 | 0.458 | 0.369 | 0.216 | 0.675 | 0.414 |
| Win32-Evul-8192-b | 1.790 | 1.789 | 1.789 | 0.495 | 0.000 | 0.003 | 0.161 | 0.036 | 0.161 | 0.179 | 0.028 | 0.475 | 0.509 | 0.324 | 0.415 | 0.328 | 0.217 | 0.600 | 0.376 |
| Win32-Evul-8192-c | 1.800 | 1.798 | 1.798 | 0.407 | 0.070 | 0.000 | 0.158 | 0.032 | 0.158 | 0.185 | 0.024 | 0.488 | 0.514 | 0.329 | 0.427 | 0.330 | 0.222 | 0.613 | 0.387 |
| Win32-Evul-8192-d | 1.926 | 1.925 | 1.925 | 0.494 | 0.831 | 0.831 | 0.000 | 0.149 | 0.000 | 0.056 | 0.153 | 0.618 | 0.654 | 0.452 | 0.566 | 0.449 | 0.303 | 0.747 | 0.543 |
| Win32-Evul-8192-e | 1.675 | 1.674 | 1.674 | 0.000 | 0.495 | 0.407 | 0.494 | 0.000 | 0.149 | 0.215 | 0.004 | 0.523 | 0.543 | 0.340 | 0.458 | 0.369 | 0.216 | 0.675 | 0.414 |
| Win32-Evul-8192-f | 1.926 | 1.925 | 1.925 | 0.494 | 0.831 | 0.831 | 0.000 | 0.494 | 0.000 | 0.056 | 0.153 | 0.618 | 0.654 | 0.452 | 0.566 | 0.449 | 0.303 | 0.747 | 0.543 |
| Win32-Evul-8192-g | 2.120 | 2.119 | 2.119 | 0.942 | 0.634 | 0.636 | 0.568 | 0.942 | 0.568 | 0.000 | 0.207 | 0.611 | 0.653 | 0.468 | 0.558 | 0.459 | 0.340 | 0.710 | 0.539 |
| Win32-Evul-8192-h | 1.742 | 1.741 | 1.741 | 0.139 | 0.379 | 0.292 | 0.627 | 0.139 | 0.627 | 0.824 | 0.000 | 0.528 | 0.555 | 0.344 | 0.463 | 0.369 | 0.227 | 0.670 | 0.418 |
| lib | 1.397 | 1.395 | 1.395 | 1.883 | 1.957 | 1.983 | 2.115 | 1.883 | 2.115 | 2.250 | 1.962 | 0.000 | 0.137 | 0.165 | 0.125 | 0.254 | 0.381 | 0.458 | 0.218 |
| write | 1.490 | 1.488 | 1.488 | 2.070 | 2.151 | 2.185 | 2.266 | 2.070 | 2.266 | 2.436 | 2.175 | 0.678 | 0.000 | 0.158 | 0.062 | 0.203 | 0.391 | 0.482 | 0.127 |
| wordconv | 1.169 | 1.169 | 1.169 | 1.482 | 1.558 | 1.587 | 1.731 | 1.482 | 1.731 | 1.915 | 1.556 | 0.946 | 0.927 | 0.000 | 0.147 | 0.214 | 0.250 | 0.372 | 0.174 |
| help | 1.469 | 1.466 | 1.466 | 1.741 | 1.797 | 1.816 | 2.009 | 1.741 | 2.009 | 2.102 | 1.785 | 0.723 | 0.298 | 0.932 | 0.000 | 0.208 | 0.342 | 0.437 | 0.151 |
| find | 1.345 | 1.343 | 1.343 | 1.538 | 1.646 | 1.671 | 1.736 | 1.538 | 1.736 | 1.927 | 1.607 | 1.101 | 0.875 | 1.118 | 0.946 | 0.000 | 0.313 | 0.598 | 0.153 |
| drvins32 | 1.719 | 1.719 | 1.719 | 1.354 | 1.468 | 1.491 | 1.563 | 1.354 | 1.563 | 1.766 | 1.449 | 1.656 | 1.673 | 1.325 | 1.561 | 1.588 | 0.000 | 0.601 | 0.318 |
| perlglob | 1.682 | 1.679 | 1.679 | 1.779 | 1.839 | 1.867 | 1.987 | 1.779 | 1.987 | 2.131 | 1.838 | 1.420 | 1.604 | 1.326 | 1.452 | 1.852 | 1.825 | 0.000 | 0.604 |
| LogonUI-til | 1.461 | 1.458 | 1.459 | 1.854 | 1.854 | 1.889 | 2.091 | 1.854 | 2.091 | 2.164 | 1.901 | 1.044 | 0.682 | 1.051 | 0.796 | 0.819 | 1.647 | 1.828 | 0.000 |

Minkowski-Form Distance (r=1)

Minkowski-Form Distance (r=2)

## VI. CONCLUSIONS AND FUTURE WORKS

This study shows that the statistical properties of opcodes can be used as a feature to detect variants of obfuscated viruses.

There are two major weaknesses in proposed method. Firstly, because a broad range of programs use some of the most usual machine instructions and this method is highly depend on instruction frequency, is very difficult to choose right threshold to decrease the risk of false positive and false negative, as well. Secondly, it works only for a limited range of obfuscation techniques. As it is obvious, for example in three mentioned cases of *Evul* virus, some metamorphic methods, such as instruction substitution and junk code insertion, can simply crush this methodology.

For the future development of this study, firstly, we recommend use of a weighted calculation of Minkowski-form distance metric. Some instructions, such as *mov*, *push*, *call*, and so on, are more employed in all programs. These kinds of opcodes can be weighted to create a more precise distance metric.

Another valuable development is to modify methodology to overcome the other obfuscation techniques that are not considered in above solution. Before we start the comparison step, the programs and their subroutines can analysed to prune garbage codes inserted or solve the issue of mutation via exchangeable instructions to obtain a uniform minimum core. No need to say that this pre-process may extremely increase the time complexity of the algorithm.

Also, normalization of the histogram will eliminate the length of frequency vector. In this study, we had to normalize the histograms to attain a threshold-based comparison for classification. However, an advantageous study is to resolve the threshold problem, eliminate the normalization of histogram, and compare the histograms in keeping with the number of opcodes, not according to proportion of frequency of opcodes.